\documentclass[runningheads]{llncs}
\usepackage[T1]{fontenc}
\usepackage{graphicx}
\usepackage{tabularray}
\usepackage{algorithm}
\usepackage{algorithmic}
\usepackage{amsmath}

\begin{document}
\title{Enhancing Speech Recognition through Text-to-Speech and Voice Conversion Augmentation}
\titlerunning{ASR Enhancement with TTS and Voice Conversion}
\author{Yunus Emre Ozkose \and
Ali Haznedaroglu}
\authorrunning{Ozkose et al.}
\institute{Sestek, Istanbul, Turkey\\
\email{\{yunusemre.ozkose,ali.haznedaroglu\}@sestek.com}}
\maketitle              % typeset the header of the contribution
\begin{abstract}
Given the challenges associated with obtaining large volumes of real-world audio, the advancement of Automatic Speech Recognition (ASR) systems increasingly depends on synthetic data. This study focuses on the call center-banking domain, providing a targeted analysis of ASR models in this context. We evaluate real and synthetic datasets, using Word Error Rate (WER) and Character Error Rate (CER) as metrics, and employ a speech quality model to assess the impact on ASR performance. Our research compares Text-to-Speech (TTS) and advanced voice conversion methods, including KNNVC, Seed-VC, and Vec2Wav2, revealing significant improvements in speech quality and ASR accuracy, particularly with Seed-VC. Additionally, our domain-specific experimentation provides insight into the unique challenges and opportunities that arise when applying ASR technologies to industry-relevant settings. This highlights the potential of voice conversion technologies to enhance ASR systems, guiding future research in diverse linguistic scenarios, and paving the way for the broader application of ASR innovations across various fields.

\end{abstract}

\section{Introduction}
Automatic Speech Recognition (ASR) systems have become integral to numerous applications, ranging from voice-activated assistants to real-time transcription services. As the demand for these systems grows, so does the need for accurate and versatile ASR models capable of understanding diverse speech patterns and dialects. Traditionally, the development of robust ASR systems relies heavily on extensive datasets of real-world audio, which capture the complexity and variability inherent in human speech. However, acquiring such datasets is often resource-intensive and may not always be feasible, especially for languages or dialects with limited speaker resources.

As an alternative, synthetic data generation has gained traction, leveraging advancements in Text-to-Speech (TTS) technologies and voice conversion methods to augment training datasets. TTS offers a straightforward approach to synthetic data generation but often falls short of replicating the nuanced attributes of natural speech, resulting in suboptimal ASR performance. In contrast, voice conversion techniques have shown promise in enhancing synthetic speech quality, with models like KNNVC, Seed-VC, and Vec2Wav2 introducing sophisticated transformations aimed at mimicking real-world audio characteristics more closely.

This study evaluates the effectiveness of these diverse data sources—real-world audio, TTS, and advanced voice conversion methods—in training ASR systems. By analyzing the performance impacts of these data types through metrics such as Word Error Rate (WER) and Character Error Rate (CER), we aim to delineate the strengths and limitations of each approach. Our findings highlight the crucial role of authentic data while demonstrating the potential of voice conversion techniques like Seed-VC to substantially improve synthetic alternatives, thus offering a pathway to more accessible ASR solutions in scenarios where real data availability is constrained.

Through this exploration, we seek to inform the broader ASR research community of the practical implications and future directions in synthetic data utilization, inspiring further innovation in the field of speech technology development.

Our study makes several novel contributions to the field of ASR:

\paragraph{Domain-Specific Experimentation:} Unlike previous research efforts that predominantly focus on the Common Voice dataset, our experiments are conducted within the specific context of the call center-banking domain. This approach allows for a more targeted analysis and evaluation of ASR models within a real-world, industry-relevant setting that presents unique challenges and opportunities.

\paragraph{Comprehensive Data Analysis:} We provide a detailed examination and comparison of real and generated audio data using a speech quality model. This analysis sheds light on the distributional characteristics and quality metrics of each dataset, thereby offering insights into the potential impact on ASR performance and the fidelity of synthetic speech generation.

\paragraph{Comparative Evaluation of Synthesis Techniques:} Our work includes a comparative analysis of Text-to-Speech (TTS) and three distinct voice cloning models. This evaluation highlights the differences in speech quality, diversity, and distribution, providing a nuanced understanding of the strengths and limitations of each approach in replicating real-world audio characteristics and improving ASR accuracy.

\section{Related Works}
Modern Text-to-Speech (TTS) synthesis has been transformed by deep learning, moving from rigid concatenative methods to flexible neural models. The dominant paradigm for a long time was a two-stage pipeline, exemplified by Tacotron 2~\cite{shen2018natural}, which first converts text into a mel-spectrogram, followed by a neural vocoder like WaveNet~\cite{oord2016wavenet} or HiFi-GAN~\cite{kong2020hifi} to generate the final waveform. To address the slow, autoregressive nature of these models, non-autoregressive systems like FastSpeech 2~\cite{ren2020fastspeech} were introduced, enabling faster inference. A significant leap in quality and efficiency came with end-to-end models like VITS~\cite{kim2021conditional}, which jointly train the text-to-spectrogram and vocoder components in a single network using variational inference and generative adversarial networks. The most recent frontier treats TTS as a language modeling problem. Models such as VALL-E~\cite{wang2023neural} and Spear-TTS~\cite{kharitonov2023speak} quantize audio into discrete tokens and use a transformer architecture to predict these tokens from text, enabling remarkably human-like synthesis and in-context learning capabilities.

Voice Cloning (VC) aims to generate speech in a specific target voice, typically from a small audio sample. This is pursued through two main paradigms. The first and most common for synthesis from text is based on multi-speaker Text-to-Speech (TTS). This approach, popularized by systems like SV2TTS~\cite{jia2018transfer}, conditions a TTS model on a speaker embedding extracted from reference audio, enabling zero-shot cloning for speakers unseen during training.

A second, parallel methodology is direct Voice Conversion, which transforms the speaker identity of a source audio utterance into that of a target speaker while preserving the linguistic content. This field includes classic non-parametric methods like kNN-VC~\cite{baas2023voiceconversionjustnearest}, which performs conversion by finding the k-nearest neighbors of source speech features within a target speaker's feature space. More recent deep learning models like AutoVC~\cite{qian2019autovc} have achieved high-quality, zero-shot conversion by using an autoencoder with a carefully designed information bottleneck to disentangle speech content from speaker identity. The ultimate goal for both paradigms is high-fidelity speaker mimicry, with the latest generation of TTS models like VALL-E~\cite{wang2023neural} blurring the lines by using reference audio as an acoustic prompt to clone not only timbre but also prosody and acoustic environment with remarkable realism.

The capacity of these systems to generate linguistically identical content across a multitude of cloned speaker identities is particularly valuable for data augmentation, as it can be used to train ASR models that are more robust to speaker variability. Li et al.~\cite{li2018training} proposed augmenting ASR training data with synthetic speech generated by a Tacotron 2-based TTS system enhanced with Global Style Tokens (GST), enabling multi-speaker modeling and prosody variation. They synthesized a complete version of the LibriSpeech training set and combined it with natural speech in a 1:1 ratio, training deep convolutional models (Wave2Letter+) with up to 54 layers. This approach led to significant WER improvements, demonstrating that large-scale synthetic augmentation with controlled speaking styles can serve as an effective regularization method and improve generalization in end-to-end ASR models. Baas and Kamper~\cite{baas2022voiceconversionimproveasr} proposed a VC framework for data augmentation aimed at extremely low-resource settings. Their architecture factorizes the speech signal by employing a pretrained CPC-based feature extractor, a quantization-based content encoder to preserve linguistic information, and a hierarchical global style token (HGST) module to capture speaker identity. They demonstrated the cross-lingual transferability of this approach, showing that a model trained exclusively on high-resource English data could generate synthetic training data for unseen languages, thereby improving the performance of fine-tuned wav2vec 2.0-based ASR systems. Casanova et al.~\cite{casanova2023asrdataaugmentationlowresource} proposed a cross-lingual data augmentation pipeline for ASR leveraging the YourTTS model, a zero-shot multi-speaker TTS and voice conversion system trained with only one speaker in the target language. By synthesizing training data using cross-lingual speaker embeddings and applying audio augmentations, they achieved over 30\% absolute WER improvement in Brazilian Portuguese and Russian, demonstrating the feasibility of training ASR models in extremely low-resource conditions. Ogun et al.~\cite{ogun2025exhaustiveevaluationttsvcbased} conducted a comprehensive analysis of synthetic data augmentation for ASR, using flow-based multi-speaker TTS and VC models to isolate the effect of augmenting specific speech attributes such as phonetic content, speaker identity, pitch, and duration. They showed that carefully controlling phoneme diversity, speaker embeddings, and environmental conditions leads to up to 35\% relative WER reduction on LibriSpeech, while naïve mixing of real and synthetic data often underperforms due to mismatched distributions.

\section{Methodology}
\subsection{Data}
The dataset most frequently utilized for Automatic Speech Recognition (ASR) tasks is the Common Voice (CV) dataset~\cite{ardila2020commonvoicemassivelymultilingualspeech}, which is publicly accessible and widely employed for ASR data augmentation studies~\cite{casanova2023asrdataaugmentationlowresource,ogun2025exhaustiveevaluationttsvcbased}. Nevertheless, there remains a significant gap in the exploration of custom domain data. In particular, this study investigates the audio domain of call centers, specifically for the Turkish language, using internally collected data. This call center data serves as a foundational benchmark for real-world data.

Access to a dataset comprising a diverse range of speakers is crucial for generating speech samples with varied vocal characteristics. For this purpose, we utilized the Librispeech dataset~\cite{panayotov2015librispeech}, which includes a total of 1281 speakers.

The initial step in creating a training dataset involves generating speech data via off-the-shelf text-to-speech (TTS) technology. We employed Azure-TTS~\cite{microsoft_azure_tts} to produce 20 hours of speech, utilizing transcriptions from authentic data. The voices "EmelNeural" and "AhmetNeural" were selected at random for this purpose. Hence, the number of speakers is initially 2. This dataset is referred to as "TTS" in the results. Subsequently, these generated speech samples were adapted to different speakers using various voice cloning models. For faster adaptation, we follow Algorithm~\ref{alg:speakerSelection}. After cloning, the number of speakers became 1281. Test data does not include synthetic data and contains 5 hours of real speech.

\begin{algorithm}[t]
\caption{Speaker Selection for TTS Voice Conversion}
\label{alg:speakerSelection}
\begin{algorithmic}[1]
\STATE \textbf{Input:} Total Speakers $N$, Audios per Speaker $M = 100$
\STATE \textbf{Output:} Converted Audios with Selected Speaker

\FOR{each unprocessed audio batch}
    \STATE Randomly select a speaker $S_i$ from the pool of $N$ speakers
    \FOR{each audio $A_j$ in the batch of $M$ audio samples}
        \STATE Convert $A_j$ to use the voice of selected speaker $S_i$
    \ENDFOR
    \STATE Log conversion details
\ENDFOR
\end{algorithmic}
\end{algorithm}

\subsection{ASR Model}
Conformer~\cite{gulati2020conformerconvolutionaugmentedtransformerspeech} is a hybrid architecture that integrates convolutional neural networks with transformers, specifically designed to excel in sequence modeling tasks like automatic speech recognition. It combines multi-head self-attention to capture long-range dependencies with convolutional layers for local feature extraction, offering a balance between global context and local detail processing. This combination allows the Conformer to effectively address the limitations of standard transformer models in handling sequential data by enhancing both temporal and spatial feature understanding.

Zipformer~\cite{yao2024zipformerfasterbetterencoder} is an adaptive architecture optimized for efficiency in computational and memory resources, tailored for real-time tasks such as speech and language processing. It features a dynamic mechanism that adjusts its layer complexity based on the input data, allowing it to manage resources effectively by compressing or expanding layers as needed. This adaptability not only improves processing speed and reduces latency but also makes the Zipformer particularly suitable for deployment in resource-constrained environments, maintaining performance while conserving computational power.

\subsection{Voice Conversion Models}
\subsubsection{k-Nearest Neighbors Voice Conversion (KNN-VC)}
~\cite{baas2023voiceconversionjustnearest} is a voice conversion framework that utilizes the simplicity and effectiveness of the k-nearest neighbors algorithm to transform speaker characteristics. The model operates by first extracting acoustic features from both the source and target voice samples, typically using methods like Mel-frequency cepstral coefficients (MFCCs). During the conversion process, KNN-VC searches for the k-nearest feature vectors in the target dataset that match the source feature vectors. The transformation is accomplished by averaging these neighbors to estimate the corresponding target features, thereby adapting the source speech to closely resemble the target speaker's attributes while preserving the original linguistic content. This approach is data-efficient as it does not require the extensive training typical of deep learning models, making it adaptable and easy to implement for real-time applications with moderate computing resources.

\subsubsection{Seed-VC}
~\cite{liu2024zeroshotvoiceconversiondiffusion} is a pioneering voice conversion system designed to generate high-quality transformations even with minimal training data, known as "seed" data. The model is constructed around advanced neural network architectures, such as GANs (Generative Adversarial Networks) or VAEs (Variational Autoencoders), which allow it to generalize from small datasets. By leveraging these sophisticated techniques, Seed-VC effectively learns to embody and reproduce the speaker-specific attributes such as pitch, timbre, and prosody of the target voice from limited input data. This is achieved through a fine-tuning process where the network parameters are adjusted to optimize the conversion quality. Seed-VC's ability to deliver personalized and convincing voice transformations using minimal samples makes it particularly advantageous in scenarios with data scarcity, enabling applications in personalization and customized voice applications across various languages and accents.

\subsubsection{vec2wav2} 
~\cite{guo2024vec2wav20advancingvoice} employs WavLM and vq-wav2vec to encode the source speech. By utilizing these technologies, the speech is encoded for both prompting and content purposes, respectively. Subsequently, the encoded speech is converted into discrete tokens, which are then softened using a Conformer encoder. This process enhances the features' flexibility and adaptability. Finally, the softened features are resynthesized into high-quality audio through the use of a BigVGAN vocoder, thereby producing output that closely resembles natural speech.

The three voice conversion models exhibit distinct approaches and strengths. KNN-VC utilizes the simplicity of the k-nearest neighbors algorithm, relying on feature matching and averaging, making it data-efficient and suitable for real-time applications without demanding significant computational resources. Seed-VC employs advanced neural architectures like GANs and VAEs, excelling in scenarios with minimal data by learning to emulate speaker-specific characteristics through sophisticated fine-tuning processes. In contrast, vec2wav2 uses technologies such as WavLM, vq-wav2vec, and BigVGAN to encode and resynthesize speech, emphasizing output quality with moderate data requirements. Each model offers unique advantages tailored to specific environments, balancing between data efficiency, adaptability, and high-quality audio output.

The criteria for selecting voice conversion models in this study are multifaceted and carefully considered to ensure optimal performance and applicability. Firstly, the models must be cross-lingual, a feature that enables them to effectively handle and convert speech across different languages. This capability is particularly important for research that involves diverse linguistic datasets, such as Turkish call center audio, which can benefit significantly from cross-lingual agility. Secondly, the models are required to exhibit state-of-the-art accuracy in speech generation. This criterion ensures that the models produce high-fidelity outputs that closely mimic the natural characteristics of human speech, thereby enhancing the realism and effectiveness of the training dataset for the ASR system. Lastly, the voice conversion models need to demonstrate proficiency with short reference speech durations. This capability is crucial for practical applications, as it allows for accurate voice conversion even when limited speech samples are available. This can be particularly advantageous in scenarios where extensive audio data are not readily accessible or where rapid conversion is desired. Together, these criteria make the selected models well-suited for the purposes of this study, ensuring both the quality and versatility of the generated speech data.

\subsubsection{Mesauring Generated Data}
We evaluate the similarity between real-world audio data and speech data generated through synthetic means. To accomplish this, we employ the NISQA framework~\cite{Mittag_2021}, which provides a robust metric for assessing the quality of both authentic and artificially generated speech data. By leveraging NISQA, we can quantitatively examine the distribution of speech quality within both data sets, thereby facilitating a detailed comparison of their respective characteristics. This approach allows us to observe not only the individual quality metrics but also the overall distribution patterns, thereby revealing pertinent differences between real and synthetic speech. Such insights are crucial for understanding the effectiveness and limitations of synthetic data in replicating the nuances of real-world speech.

\section{Experiments}
Initially, the specified ASR model is trained using the real-world dataset with varying durations of audio data. Following this phase, the training dataset is substituted with speech data generated through TTS and VC models.

\subsubsection{Speech Qualities} 
\begin{figure}[t]
\includegraphics[width=\textwidth]{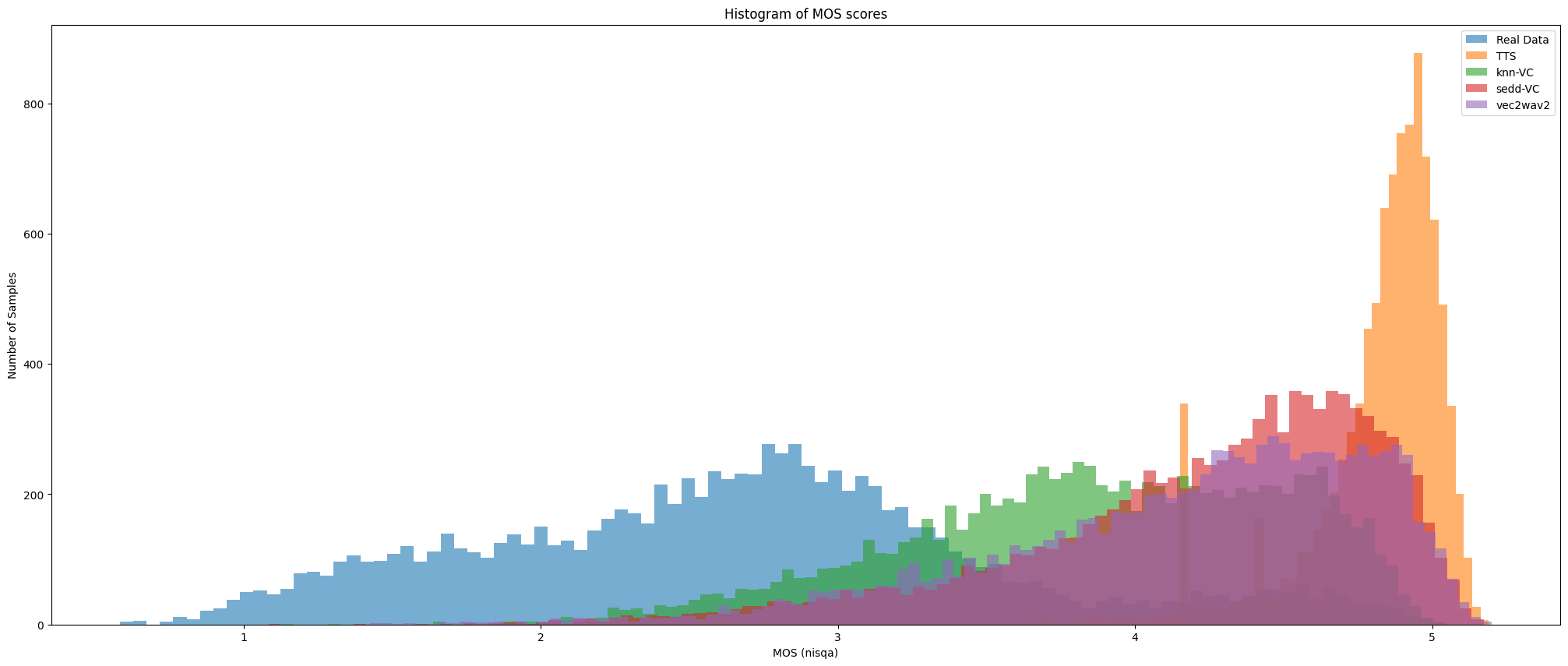}
\caption{Histogram of NISQA-predicted MOS (Mean Opinion Score) values for real and synthetic speech data. The "Real Data" (blue) shows a broad distribution of perceived quality, reflecting diverse acoustic conditions. The TTS system (orange) yields consistently high scores with a sharp peak near 5.0, indicating very high synthetic quality but limited variability. Voice conversion methods (knn-VC in green, sedd-VC in red, and vec2wav2 in purple) produce more naturally distributed scores, with sedd-VC and vec2wav2 achieving high MOS values while preserving greater variability than TTS. These results highlight the trade-off between synthetic clarity and realism across systems.} \label{fig1}
\end{figure}

The comparative distributions of real and generated data qualities are illustrated in Fig.~\ref{fig1}. The real-world audio data exhibit greater variability and a higher level of noise compared to the generated speech samples, which display significantly less diversity. The Text-to-Speech (TTS) generated speeches tend to cluster within a high-quality range. However, their distribution does not align with that of the real data, indicating a disparity in variability and naturalness.

Conversely, the speech samples produced through cloning techniques exhibit a distribution that more closely resembles the variability found in real-world data. By converting TTS-generated speech to match the vocal characteristics of speakers from the Librispeech dataset, the resulting data distribution becomes more similar to the authentic data. This observation suggests that voice conversion techniques enhance the representativeness and variability of synthetic speech, thereby bridging the gap between generated and real audio data.

Consequently, this finding supports the hypothesis that voice conversion can be more effective than TTS alone in aiding speech recognition systems to reduce word error rates. The improved alignment of the cloned speech distribution with real data exemplifies its potential to enhance the robustness and accuracy of ASR systems in practical applications.

\subsubsection{Experimental Setup for ASR Model}
ASR model is inially trained on english data, and then finetuned with the specified cases. Byte Pair Encoding (BPE) is employed for tokenization, utilizing a vocabulary size of 500 tokens. The context size is configured to 2. For performance evaluation, we conduct an averaging of the final epochs, specifically from the latest 1st, 3rd, 5th, 7th, and 9th epochs, and report the optimal result obtained. We use K2\footnote{https://github.com/k2-fsa/k2} and Icefall\footnote{https://github.com/k2-fsa/icefall} tools for training.

\subsection{Real Data and TTS}
The results in Table \ref{tab1} provide a comparison of Word Error Rate (WER) and Character Error Rate (CER) for different datasets used in training an Automatic Speech Recognition (ASR) system over a 20-hour period. As expected, the real data outperforms synthetic methods, achieving a WER of 34.40\% and a CER of 15.33\%. This highlights the effectiveness of real-world audio in capturing the nuances and variability inherent in natural speech, which is vital for developing robust ASR models.

Text-to-Speech (TTS) stands out with a notably high WER of 97.10\% and a CER of 80.18\%, suggesting significant limitations in replicating the natural qualities of human speech required for accurate recognition. The synthetic speech generated by TTS lacks the variations and natural characteristics, leading to a reduced performance.

KNNVC, a voice conversion method, achieves a WER of 83.06\% and a CER of 55.08\%. This technique surpasses TTS in performance, indicating its ability to generate more natural-sounding synthetic speech, yet it still trails substantially behind real data. Seed-VC yields better results, with a WER of 75.53\% and a CER of 43.97\%. The relative improvement in WER over TTS is approximately 22.2\%, suggesting Seed-VC's superior capability in synthesizing speech that faithfully mimics the qualities of real-world audio.

Vec2Wav2 demonstrates moderate effectiveness with a WER of 78.42\% and a CER of 46.62\%, providing improvements over both TTS and KNNVC but slightly underperforming compared to Seed-VC. The relative improvement in WER compared to TTS is approximately 19.2\%, highlighting Vec2Wav2's potential in enhancing synthetic speech for ASR training. Overall, these findings indicate that while real audio data remains crucial for optimal ASR system performance, advanced voice conversion techniques, particularly Seed-VC, can significantly boost the utility of synthetic datasets in speech recognition applications.

\begin{table}[t]
\caption{Word Error Rate (WER) comparison for different types of training data in an Automatic Speech Recognition (ASR) system, evaluated over training duration of 20 hours. The results highlight the effectiveness of real-world audio data and various synthetic speech generation methods, including TTS and voice conversion techniques such as KNNVC, Seed-VC, and Vec2Wav2.}\label{tab1}
\begin{tblr}{|l|X[c]|X[c]|}
\hline
\textbf{Train Data} & \textbf{CER $\downarrow$} & \textbf{WER $\downarrow$} \\
\hline
Real Data & 15.33 & 34.40 \\
TTS & 80.18 & 97.10 \\
KNNVC & 55.08 & 83.06  \\
Seed-VC & 43.97 & 75.53 \\
Vec2Wav2 & 46.62 & 78.42 \\
\hline
\end{tblr}
\end{table}

\subsection{Substituting Real Speech Data with Synthetic Clones via Voice Conversion Models}
The presented results in Table \ref{tab2} provide an insightful examination of Word Error Rate (WER) and Character Error Rate (CER) when utilizing enhanced voice conversion models and their combinations with Text-to-Speech (TTS) in training Automatic Speech Recognition (ASR) systems over a 20-hour duration. Compared to our previous discussion tables, these configurations demonstrate the effectiveness of repeated applications of KNNVC, Seed-VC, and Vec2Wav2 voice conversion methods, both independently and integrated with TTS-generated speech.

Voice conversion techniques alone display varying levels of effectiveness when applied repeatedly. KNNVC applied five times results in the highest WER of 82.25\%, indicating a significant error rate that reflects potential overfitting problems or artifacts introduced through excessive synthesis transformations. This suggests that without the refinement of TTS integration, pure voice conversion might lack the necessary naturalness or variability found in real-world speech. 

Seed-VC applied five times achieves a notable reduction in WER to 73.99\%, representing a relative improvement of approximately 9.92\% compared to KNNVC. This suggests that Seed-VC's repeated transformations produce synthetic speech that is closer in acoustic features to real speech, thereby improving ASR model performance.

Vec2Wav2 repeated application, resulting in a WER of 75.77\%, offers a more moderate improvement compared to KNNVC, with a relative reduction of around 7.86\%. This demonstrates that while Vec2Wav2 improves performance, it yields less pronounced gains than Seed-VC in its fivefold application, highlighting the specific strengths and limitations inherent in each voice conversion model.

The integration of TTS with voice conversion techniques reveals additional enhancements. The combination of TTS with KNNVC leads to a slight reduction in WER to 81.8\%, yielding a minor relative improvement of approximately 0.55\% over KNNVC alone. However, the benefits are more pronounced when TTS is combined with Seed-VC, where WER reduces further to 74.62\%, a relative improvement of about 9.33\% compared to TTS with KNNVC, underscoring the complementary nature of Seed-VC and TTS.

Similarly, while combining TTS with Vec2Wav2 results in a WER of 76.82\%, this configuration also provides a small yet notable relative improvement of around 5.48\% compared to Vec2Wav2 alone. This indicates that TTS can enhance the performance of voice conversion techniques, though the extent varies depending on the particular model.

Overall, the results suggest that voice conversion models, particularly when integrated with TTS, can improve speech recognition accuracy. Nonetheless, the degree of improvement is contingent on the model and approach used, with Seed-VC demonstrating the most substantial gains, especially when paired with TTS. These findings highlight the critical role of strategic integration and moderation in applying voice conversion and TTS techniques to optimize ASR systems' performance effectively.

\begin{table}[t]
\caption{Word Error Rate (WER) analysis of speech recognition performance using enhanced voice conversion models and combinations with Text-to-Speech (TTS). Evaluated over training duration of 20 hours, this table presents results for models trained on synthetic speech generated through repeated applications of KNNVC, Seed-VC, and Vec2Wav2, as well as the integration of TTS-generated speech with these voice conversion techniques.}\label{tab2}
\begin{tblr}{|l|X[c]|X[c]|}
\hline
\textbf{Train Data} & \textbf{CER $\downarrow$} & \textbf{WER $\downarrow$} \\
\hline
KNNVC $\times$5 & 52.97 & 82.25  \\
Seed-VC $\times$5 & 41.47 & 73.99  \\
Vec2Wav2 $\times$5 & 44.4 & 75.77  \\
\hline
TTS + KNNVC & 52.19 & 81.8  \\
TTS + Seed-VC & 43.95 & 74.62  \\
TTS + Vec2Wav2 & 46.42 & 76.82  \\
\hline
\end{tblr}
\end{table}

\subsection{Augmenting Real Speech Data with Synthetic Clones}
The results in Table \ref{tab3} illustrate the effectiveness of different combinations of TTS and voice conversion techniques in reducing the Word Error Rate (WER) in Automatic Speech Recognition (ASR) systems with 20 hours of training data. Each configuration demonstrates a considerable improvement over the assumed baseline WER of 34.40\% obtained with real data alone.

Integrating TTS with the Vec2Wav2 model yields the most significant reduction, achieving a WER of 28.65\%. This translates to a relative improvement of approximately 16.74\% compared to the baseline, highlighting the efficacy of this combination in enhancing speech recognition accuracy. However, applying Vec2Wav2 fivefold increases the WER to 30.26\%, which still represents a relative improvement of about 12.04\% but points to potential over-processing issues.

Similarly, the TTS plus KNNVC configuration results in a WER of 28.98\%, corresponding to a relative improvement of approximately 15.78\%. Yet, when KNNVC is applied five times, the WER rises to 30.26\%, reducing the relative improvement to roughly 12.04\%, consistent with findings for multiple Vec2Wav2 applications.

The integration of TTS with Seed-VC results in a WER of 32.30\%, yielding a modest improvement of about 6.10\%. However, applying Seed-VC fivefold provides a notable WER of 29.71\%, improving the relative figure to 13.63\%, suggesting that multiple applications can be beneficial for Seed-VC, contrary to trends observed with other techniques.

Finally, using all clones with TTS results in a WER of 29.01\%, representing a relative improvement of approximately 15.67\%. This balanced approach delivers robust performance without the complexities associated with multiple applications of a single technique.

In summary, these results clearly demonstrate that while all configurations enhance WER over the baseline, the degree of improvement is variable. The Vec2Wav2 technique, notably in its single-application form, produces the most substantial relative gains. Importantly, the analysis suggests that applying voice conversion techniques excessively can lead to less pronounced improvements, potentially due to overfitting or introduction of recognition-inhibiting artifacts. Consequently, selecting the optimal number of applications is crucial for maximizing the benefits of data augmentation in ASR systems.

% Another discussion is whether combining generated speeches that are produced by different voice conversion models helps to decrease the word error rate. ...

\begin{table}[t]
\caption{Performance comparison of Word Error Rate (WER) when blending real and synthetic training data in Automatic Speech Recognition (ASR) systems. The table highlights WER improvements with real data alone and in combination with TTS and multiple iterations of voice conversion techniques (KNNVC, Seed-VC, Vec2Wav2). Evaluated over 20 hours of training data, configurations include single and multiple applications (denoted as $\times$5) of voice conversion methods}\label{tab3}
\begin{tblr}{|l|X[c]|X[c]|}
\hline
\textbf{Train Data} & \textbf{CER $\downarrow$} & \textbf{WER $\downarrow$} \\
\hline
% Real Data & 15.33 & 34.40 \\
% \hline
Real Data + TTS + KNNVC & 12.32 & 28.98  \\
Real Data + TTS + KNNVC $\times$5 & 13.71 & 30.26  \\

Real Data + TTS + Seed-VC & 13.71 & 32.30  \\
Real Data + TTS + Seed-VC$\times$5 & 13.25 & 29.71 \\

Real Data + TTS + Vec2Wav2 & \textbf{12.12} & \textbf{28.65} \\
Real Data + TTS + Vec2Wav2 $\times$5 & 13.51 & 30.26  \\
% \hline
Real Data + TTS + all clones & 12.58 & 29.01 \\
\hline
\end{tblr}
\end{table}

\subsection{Exploring the Impact of Varying Training Durations on Model Performance}
The analysis of the Word Error Rate (WER) and Character Error Rate (CER) across different training durations (10, 20, and 50 hours) highlights several key observations on the performance enhancements achieved through data augmentation techniques, specifically TTS and various voice conversion models such as KNN-VC, Seed-VC, and Vec2Wav2, in the context of Turkish call center data.

Initially examining the WER of models trained solely on real data, a clear and substantial decrease is observed as the training duration increases: from 58.71\% with 10 hours of training to 34.40\% with 20 hours, and further to 25.48\% with 50 hours. This expected improvement underscores the importance of extensive training datasets in enhancing speech recognition models.

When comparing models augmented with TTS and voice conversion techniques to those trained only on real data, significant improvements are observed. With 10 hours of training, the integration of TTS and KNN-VC, Seed-VC, and Vec2Wav2 results in relative WER reductions of approximately 19\% to 21\%. Specifically, the WER decreases from 58.71\% to about 46-47\%, depending on the specific voice conversion model used. At 20 hours, the relative improvement in WER is between 6\% and 17\%, with the augmented models achieving WERs ranging from 28.65\% (Vec2Wav2) to 32.30\% (Seed-VC), notable reductions from the baseline of 34.40\%.

Interestingly, as the amount of training data increases to 50 hours, the relative improvements in WER obtained from the voice conversion models become less pronounced, with improvements around 1\% to 3\% when compared to models trained on real data alone. All augmented models achieve WERs around 25\%, closely aligning with the baseline and with one another, indicating diminishing returns on WER improvements from these specific techniques as training duration grows.

These findings reveal that while TTS and voice conversion models initially provide substantial advantages, particularly with limited data, their relative benefit decreases as the training datasets expand. This suggests that the effectiveness of these augmentation techniques is more pronounced in resource-constrained scenarios. Overall, the results underscore the utility of data augmentation in reducing WER and CER for Turkish call center data, emphasizing their role in enhancing model robustness and accuracy, especially in the early stages of model development.

\begin{table}[t]
\caption{An In-Depth Analysis of Character Error Rate (CER) and Word Error Rate (WER) for Various Voice Conversion Models, Evaluating Performance Across Different Training Durations: A Comprehensive Comparison for 10, 20, and 50 Hours of Training on Real Data, Augmented with TTS and Different Voice Conversion Techniques Including KNN-VC, Seed-VC, and Vec2Wav2.}\label{tab3}
\begin{tblr}{|l|X[c]|X[c]|X[c]|X[c]|X[c]|X[c]|}
\hline
 & \SetCell[c=6]{c}{\textbf{Amount of Hours in Training}} \\
 & \SetCell[c=2]{c}{{{\textbf{10 hours}}}} & & \SetCell[c=2]{c}{{{\textbf{20 hours}}}} & & \SetCell[c=2]{c}{{{\textbf{50 hours}}}} & \\
\textbf{Train Data} & \textbf{CER} & \textbf{WER}  & \textbf{CER}  & \textbf{WER} & \textbf{CER}  & \textbf{WER}  \\
\hline
Real Data & 35.57 & 58.71 & 15.33 & 34.40 & 10.44 & 25.48 \\
\hline
Real Data + TTS + KNNVC & 24.38 & 47.08 & 12.32 & 28.98 & 10.18 & 25.47 \\
Real Data + TTS + Seed-VC & 23.89 & 46.56 & 13.71 & 32.30 & 10.19 & 25.19  \\
Real Data + TTS + Vec2Wav2 & 24.5 & 46.72 & 12.12 & 28.65 & 10.04 & 25.26 \\
\hline
\end{tblr}
\end{table}

% training the model on only synthetic data tends to overfit in earlier epochs. 

\section{Conclusion}
% Jointly augmenting all other attributes reduces the WER of a Conformer-Transducer model by 11\% relative on Common Voice and by up to 35\% relative on LibriSpeech compared to training on real data only.

This study investigates the efficacy of various data sources for training Automatic Speech Recognition (ASR) systems, focusing on real-world audio, Text-to-Speech (TTS), and advanced voice conversion techniques such as KNNVC, Seed-VC, and Vec2Wav2. Our findings unequivocally establish that real-world audio data yields superior ASR performance, achieving the lowest Word Error Rate (WER) and Character Error Rate (CER). This underscores the critical role of authentic speech in capturing the nuanced variability essential for robust ASR models.

While synthetic methods offer a feasible alternative in the absence of extensive real data, their effectiveness varies significantly. The TTS approach, though widely accessible, demonstrated substantial limitations, as evidenced by its higher WER and CER, likely due to inadequate representation of natural speech dynamics. 

Conversely, the integration of voice conversion techniques presented marked improvements. Seed-VC emerged as the most promising method among the synthetic approaches, delivering the highest reduction in WER and CER relative to TTS, thereby demonstrating its potential to bridge the performance gap between synthetic and real data. Vec2Wav2 also showed efficacy, albeit to a lesser extent. These results suggest that advanced voice conversion can enhance the realism and utility of synthetic datasets for ASR training.

In conclusion, while real-world audio remains the benchmark for optimal ASR performance, the strategic implementation of sophisticated voice conversion methods presents a viable path to enhancing ASR systems, particularly in resource-constrained environments. Future research could further refine these synthetic methods and explore their broader application in diverse linguistic contexts, paving the way for more accessible and effective ASR technologies.

\bibliographystyle{splncs04}
\bibliography{bibliography.bib}

@misc{ogun2025exhaustiveevaluationttsvcbased,
      title={An Exhaustive Evaluation of TTS- and VC-based Data Augmentation for ASR}, 
      author={Sewade Ogun and Vincent Colotte and Emmanuel Vincent},
      year={2025},
      eprint={2503.08954},
      archivePrefix={arXiv},
      primaryClass={eess.AS},
      url={https://arxiv.org/abs/2503.08954}, 
}

@misc{casanova2023asrdataaugmentationlowresource,
      title={ASR data augmentation in low-resource settings using cross-lingual multi-speaker TTS and cross-lingual voice conversion}, 
      author={Edresson Casanova and Christopher Shulby and Alexander Korolev and Arnaldo Candido Junior and Anderson da Silva Soares and Sandra Aluísio and Moacir Antonelli Ponti},
      year={2023},
      eprint={2204.00618},
      archivePrefix={arXiv},
      primaryClass={eess.AS},
      url={https://arxiv.org/abs/2204.00618}, 
}

@misc{baas2023voiceconversionjustnearest,
      title={Voice Conversion With Just Nearest Neighbors}, 
      author={Matthew Baas and Benjamin van Niekerk and Herman Kamper},
      year={2023},
      eprint={2305.18975},
      archivePrefix={arXiv},
      primaryClass={eess.AS},
      url={https://arxiv.org/abs/2305.18975}, 
}

@misc{liu2024zeroshotvoiceconversiondiffusion,
      title={Zero-shot Voice Conversion with Diffusion Transformers}, 
      author={Songting Liu},
      year={2024},
      eprint={2411.09943},
      archivePrefix={arXiv},
      primaryClass={cs.SD},
      url={https://arxiv.org/abs/2411.09943}, 
}

@misc{guo2024vec2wav20advancingvoice,
      title={vec2wav 2.0: Advancing Voice Conversion via Discrete Token Vocoders}, 
      author={Yiwei Guo and Zhihan Li and Junjie Li and Chenpeng Du and Hankun Wang and Shuai Wang and Xie Chen and Kai Yu},
      year={2024},
      eprint={2409.01995},
      archivePrefix={arXiv},
      primaryClass={eess.AS},
      url={https://arxiv.org/abs/2409.01995}, 
}

@misc{ardila2020commonvoicemassivelymultilingualspeech,
      title={Common Voice: A Massively-Multilingual Speech Corpus}, 
      author={Rosana Ardila and Megan Branson and Kelly Davis and Michael Henretty and Michael Kohler and Josh Meyer and Reuben Morais and Lindsay Saunders and Francis M. Tyers and Gregor Weber},
      year={2020},
      eprint={1912.06670},
      archivePrefix={arXiv},
      primaryClass={cs.CL},
      url={https://arxiv.org/abs/1912.06670}, 
}

@misc{yao2024zipformerfasterbetterencoder,
      title={Zipformer: A faster and better encoder for automatic speech recognition}, 
      author={Zengwei Yao and Liyong Guo and Xiaoyu Yang and Wei Kang and Fangjun Kuang and Yifan Yang and Zengrui Jin and Long Lin and Daniel Povey},
      year={2024},
      eprint={2310.11230},
      archivePrefix={arXiv},
      primaryClass={eess.AS},
      url={https://arxiv.org/abs/2310.11230}, 
}

@misc{gulati2020conformerconvolutionaugmentedtransformerspeech,
      title={Conformer: Convolution-augmented Transformer for Speech Recognition}, 
      author={Anmol Gulati and James Qin and Chung-Cheng Chiu and Niki Parmar and Yu Zhang and Jiahui Yu and Wei Han and Shibo Wang and Zhengdong Zhang and Yonghui Wu and Ruoming Pang},
      year={2020},
      eprint={2005.08100},
      archivePrefix={arXiv},
      primaryClass={eess.AS},
      url={https://arxiv.org/abs/2005.08100}, 
}

@inproceedings{panayotov2015librispeech,
  title={Librispeech: an asr corpus based on public domain audio books},
  author={Panayotov, Vassil and Chen, Guoguo and Povey, Daniel and Khudanpur, Sanjeev},
  booktitle={2015 IEEE international conference on acoustics, speech and signal processing (ICASSP)},
  pages={5206--5210},
  year={2015},
  organization={IEEE}
}

@misc{microsoft_azure_tts,
  author       = {Microsoft},
  title        = {Azure Text-to-Speech},
  year         = {2023},
  howpublished = {\url{https://azure.microsoft.com/services/cognitive-services/text-to-speech/}},
  note         = {Accessed: 05, 13, 2025}
}

@inproceedings{baas2022voiceconversionimproveasr,
  title={Voice Conversion Can Improve ASR in Very Low-Resource Settings},
  author={Matthew Baas and Herman Kamper},
  booktitle={Interspeech},
  year={2021},
  url={https://api.semanticscholar.org/CorpusID:242757430}
}

@inproceedings{Mittag_2021,
   title={NISQA: A Deep CNN-Self-Attention Model for Multidimensional Speech Quality Prediction with Crowdsourced Datasets},
   url={http://dx.doi.org/10.21437/Interspeech.2021-299},
   DOI={10.21437/interspeech.2021-299},
   booktitle={Interspeech 2021},
   publisher={ISCA},
   author={Mittag, Gabriel and Naderi, Babak and Chehadi, Assmaa and Möller, Sebastian},
   year={2021},
   collection={interspeech_2021} 
}

@article{li2018training,
  title={Training neural speech recognition systems with synthetic speech augmentation},
  author={Li, Jason and Gadde, Ravi and Ginsburg, Boris and Lavrukhin, Vitaly},
  journal={arXiv preprint arXiv:1811.00707},
  year={2018}
}

@inproceedings{shen2018natural,
  title={Natural tts synthesis by conditioning wavenet on mel spectrogram predictions},
  author={Shen, Jonathan and Pang, Ruoming and Weiss, Ron J and Schuster, Mike and Jaitly, Navdeep and Yang, Zongheng and Chen, Zhifeng and Zhang, Yu and Wang, Yuxuan and Skerrv-Ryan, Rj and others},
  booktitle={2018 IEEE international conference on acoustics, speech and signal processing (ICASSP)},
  pages={4779--4783},
  year={2018},
  organization={IEEE}
}

@article{oord2016wavenet,
  title={Wavenet: A generative model for raw audio},
  author={Oord, Aaron van den and Dieleman, Sander and Zen, Heiga and Simonyan, Karen and Vinyals, Oriol and Graves, Alex and Kalchbrenner, Nal and Senior, Andrew and Kavukcuoglu, Koray},
  journal={arXiv preprint arXiv:1609.03499},
  year={2016}
}

@article{kong2020hifi,
  title={Hifi-gan: Generative adversarial networks for efficient and high fidelity speech synthesis},
  author={Kong, Jungil and Kim, Jaehyeon and Bae, Jaekyoung},
  journal={Advances in neural information processing systems},
  volume={33},
  pages={17022--17033},
  year={2020}
}

@article{ren2020fastspeech,
  title={Fastspeech 2: Fast and high-quality end-to-end text to speech},
  author={Ren, Yi and Hu, Chenxu and Tan, Xu and Qin, Tao and Zhao, Sheng and Zhao, Zhou and Liu, Tie-Yan},
  journal={arXiv preprint arXiv:2006.04558},
  year={2020}
}

@inproceedings{kim2021conditional,
  title={Conditional variational autoencoder with adversarial learning for end-to-end text-to-speech},
  author={Kim, Jaehyeon and Kong, Jungil and Son, Juhee},
  booktitle={International Conference on Machine Learning},
  pages={5530--5540},
  year={2021},
  organization={PMLR}
}

@article{wang2023neural,
  title={Neural codec language models are zero-shot text to speech synthesizers},
  author={Wang, Chengyi and Chen, Sanyuan and Wu, Yu and Zhang, Ziqiang and Zhou, Long and Liu, Shujie and Chen, Zhuo and Liu, Yanqing and Wang, Huaming and Li, Jinyu and others},
  journal={arXiv preprint arXiv:2301.02111},
  year={2023}
}

@article{kharitonov2023speak,
  title={Speak, read and prompt: High-fidelity text-to-speech with minimal supervision},
  author={Kharitonov, Eugene and Vincent, Damien and Borsos, Zal{\'a}n and Marinier, Rapha{\"e}l and Girgin, Sertan and Pietquin, Olivier and Sharifi, Matt and Tagliasacchi, Marco and Zeghidour, Neil},
  journal={Transactions of the Association for Computational Linguistics},
  volume={11},
  pages={1703--1718},
  year={2023},
  publisher={MIT Press One Broadway, 12th Floor, Cambridge, Massachusetts 02142, USA~…}
}

@article{jia2018transfer,
  title={Transfer learning from speaker verification to multispeaker text-to-speech synthesis},
  author={Jia, Ye and Zhang, Yu and Weiss, Ron and Wang, Quan and Shen, Jonathan and Ren, Fei and Nguyen, Patrick and Pang, Ruoming and Lopez Moreno, Ignacio and Wu, Yonghui and others},
  journal={Advances in neural information processing systems},
  volume={31},
  year={2018}
}

@inproceedings{qian2019autovc,
  title={Autovc: Zero-shot voice style transfer with only autoencoder loss},
  author={Qian, Kaizhi and Zhang, Yang and Chang, Shiyu and Yang, Xuesong and Hasegawa-Johnson, Mark},
  booktitle={International Conference on Machine Learning},
  pages={5210--5219},
  year={2019},
  organization={PMLR}
}

\end{document}